\documentclass[11pt]{asaproc}

\usepackage{graphicx}
\usepackage[utf8]{inputenc}
\graphicspath{{images/}}
\usepackage{natbib}

\usepackage{times}
\usepackage[algo2e]{algorithm2e}
\usepackage{algorithm}
\usepackage{mathtools}
\usepackage{algorithmic}
\usepackage{multirow}
\usepackage{makecell}
\usepackage{url}

\DeclarePairedDelimiter\floor{\lfloor}{\rfloor}
\newcommand{\indep}{\rotatebox[origin=c]{90}{$\models$}}

\title{ Assign Experiment Variants at Scale in Online Controlled Experiments}

\author{Qike Li\thanks{Wish,
  San Francisco, CA 94104, qike.max.li@gmail.com} \and  Samir Jamkhande \thanks{Wish,
  San Francisco, CA 94104, sjamkhande@wish.com} \and  Pavel  Kochetkov\thanks{Wish,
  San Francisco, CA 94104, pkochetkov@wish.com} \and Pai Liu \thanks{Wish,
  San Francisco, CA 94104, pliu@wish.com}}
\begin{document}

\maketitle

\begin{abstract}
Online controlled experiments (A/B tests) have become the gold standard for learning the impact of new product features in technology companies. Randomization enables the inference of causality from an A/B test. The randomized assignment maps end users to experiment buckets and balances user characteristics between the groups. Therefore, experiments can attribute any outcome differences between the experiment groups to the product feature under experiment. Technology companies run A/B tests at scale – hundreds if not thousands of A/B tests concurrently, each with millions of users. The large scale poses unique challenges to randomization. First, the randomized assignment must be fast since the experiment service receives hundreds of thousands of queries per second. Second, the variant assignments must be independent between experiments. Third, the assignment must be consistent when users revisit or an experiment enrolls more users. We present a novel assignment algorithm and statistical tests to validate the randomized assignments. Our results demonstrate that not only is this algorithm computationally fast but also satisfies the statistical requirements: unbiased and independent.
\end{abstract}

\section{Introduction\label{intro}}
Online controlled experimentation, also known as A/B testing \citep{box2005statistics,kohavi2012online}, has become the gold standard for measuring the impact of new product features (e.g., a new user interface). Many companies, including Facebook, Microsoft, Google, LinkedIn, Amazon, Airbnb, Wish, and many others, use A/B testing for product development and enhancement \citep{tang2014experimentation,xu2015infrastructure,bakshy2014designing,li_2021}. 

Randomization enables inference of causality from A/B test and, therefore, is critical for user assignment. A/B tests randomly assign users to experiment groups independent of their characteristics. The randomized assignment results in balanced user characteristics, both observed and unobserved, between groups. In consequence, any outcome differences between the experiment groups (e.g., control vs. treatment) are due to the product feature under experiment.

The randomized assignment has unique challenges when technology companies run A/B tests at scale. First, the user assignment algorithm needs to be fast. The user assignment service is called hundreds of thousands of times per second to output user assignments for millions of users from hundreds of experiments in real time. Second, user assignments need to be independent between experiments. At any time, a user can be in hundreds if not thousands of A/B tests. The user assignment in any experiment should not affect his/her assignment in any other experiments. Third, the randomized assignment need to be consistent for each user. Users come to the website repeatedly, consistent assignment ensures the same user experience.

When we ran an A/A test, also known as null test \citep{kohavi2009controlled}, we discovered that our original assignment algorithm led to sample ratio mismatch ~\citep{fabijan2019diagnosing}, which refers to the mismatch between the sample ratio set by the experimenter and the observed sample ratio. Further, the original assignment algorithm also leads to correlated assignments between experiments. Namely, the probability of a user being assigned to, for example, control bucket is higher than expected if the user is in the control bucket of another experiment. 

The challenges in randomized assignment are ubiquitous in online experiments ~\citep{zhao2016online,xu2015infrastructure,macmillan_2018}. A good randomization algorithm is critical to ensure statistical validity of experiments. The algorithm must generate assignments that are unbiased towards any experiment variant, independent between experiments, and deterministic based on the combination of user ID and experiment ID.

In this paper, we present a novel assignment algorithm and demonstrate statistical tests to validate the randomized assignments.  

\section{Methodology}
\subsection{Existing Methodologies}
Assignment algorithms randomly map user IDs to experiment groups. A/B tests typically employ hash functions to transform user IDs into uniform random numbers. Facebook uses hashing function sha1 \citep{bakshy2014designing}, and Yahoo uses hashing function Fowler–Noll–Vo (FNV) \citep{fowler1994fnv} and message-digest algorithm (MD5) \citep{rivest1992md5} to transform user IDs \citep{zhao2016online}. Hash functions are deterministic and thus ensure a consistent user experience for each user in an experiment. The output of hash functions is random. The randomness ensures user characteristics are well balanced between experiment buckets and, therefore, we can draw causal conclusions from A/B tests. The output of hash functions follows a uniform distribution, which guarantees the users in an experiment have equal chances to see each variant. 

The original assignment algorithm at Wish is a two-step approach. Both steps apply the hash function FNV to map user IDs plus a suffix string (the suffixes are different in the two steps) to random numbers, which are then transformed to the range of [0,99] by taking a module of 100. The first integer $R_e$ determines if a user should be exposed to the experiment, and the second integer determines which experiment group the user is assigned to. For example, when set exposure rate at 50\% and allocate 20\% users to control and 80\% users to treatment, the experiment assigns a user with $R_e<50$ and $R_b<20$ to control and assigns a user with $R_e<50$ and $R_b\geq 20$ to treatment. Users with $R_e\geq50$ will not be exposed to the experiment. 

\begin{algorithm}[H]
\floatname{algorithm}{Original randomization algorithm}
\renewcommand{\thealgorithm}{}
\renewcommand{\listalgorithmname}{List of Algorithms}
\caption{}
\label{old-algo}
\begin{algorithmic}[1]
\STATE \textbf{Step 1}
\STATE $S^e \gets Concatenate$  (salt, user ID, "Exposure")
\STATE $H^e \gets FNVHash(S^e)$
\STATE $R^e \gets H^e \mod 100$

  \IF{$R^e > \text{exposure\_rate} \cdot 100$}
    \STATE \RETURN ignore
  \ELSE
    \STATE Go to Step 2
  \ENDIF
\STATE \textbf{Step 2}
\STATE $S^b \gets Concatenate$  (salt, user ID, "Bucket")
\STATE $H^b \gets FNVHash(S^b)$
\STATE $R^b \gets H^b \mod 100$  
 \IF{$R^b < \text{control\_bucket\_percentage} \cdot 100 $}
    \STATE \RETURN control
  \ELSE
    \STATE \RETURN treatment
  \ENDIF
\end{algorithmic}
\end{algorithm} 

\begin{figure}[htbp]
    \centering
    \includegraphics[width=10cm]{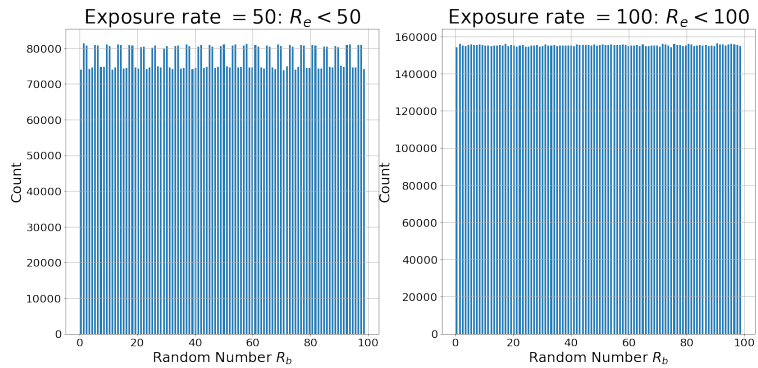}
    \caption{Histogram of $R_b|R_e<50$ and $R_b|R_e<100$}
    \label{fig:uniform}
\end{figure}

Our seemingly straightforward assignment algorithm led to non-uniform distribution of $R_b|R_e$ at some exposure rates due to the dependence between $R_e$ and $R_b$. Figure \ref{fig:uniform} compares the distribution of $R_b|R_e<50$ (left) and the marginal distribution of $R_b$ (right).

Further, the algorithm introduces dependence between experiments. That is, a user is more likely to be in, for example, a control bucket if the user is in control from another experiment. Namely, $R_b^{i} \not\!\perp\!\!\!\perp R_b^{j}$, where $i$ and $j$ represent experiment i and experiment j respectively. We present the results of independence tests in section \ref{sec:ind}.

\subsection{Proposed Methodology}

The proposed methodology aims to achieve the following goals. First, the resulting distribution of $R_b|R_e$ must be uniform, such that the algorithm assigns users to different buckets with equal probabilities. Second,  $R_b^{i} \indep R_b^{j} \quad \forall i, j$. Namely, a user's assignment in one experiment must not affect that user's assignment in another experiment. Third, consistent assignment so that a user is assigned to the same variant on successive visits. Lastly, experiment ramp-up does not change assignments of the previously assigned users.

Given these requirements, we designed our algorithm as follows. When a user visits the app, the experimentation service conducts the following steps
\begin{enumerate}

\item Concatenate experiment ID and user ID to form string $S$.
\item Apply a 64 bit hash function to map the concatenated string, $S$, to a hash value $H$. 
\item Divide $H$ by float(0xFFFFFFFFFFFFFFFF) and multiply it by 10,000 to get a uniform random number integer $Z$ ranging from 0 to 9,999.
\item Compute $R_e$ as $\frac{Z}{100}$. If $R_e \geq E\% \cdot 100$, where $E\%$ is the exposure rate (e.g., 10\%) , we assign 'ignore' to this user. An 'ignore' user will not be exposed to the experiment and will be excluded in any calculations for this experiment.

\item Compute $R_b$ as $Z \bmod 100$. When there are two experiment buckets: control, treatment, assign a user to control if  $R_b< C\%\cdot 100$, where $C\%$ is the control bucket percentage. Otherwise, assign treatment. 
\end{enumerate}

\begin{algorithm}[H]
\floatname{algorithm}{New Assignment Algorithm}
\renewcommand{\thealgorithm}{}
\caption{}
\label{new-algo}
\begin{algorithmic}[1]
\STATE $S \gets Concatenate$  (experiment ID, user ID) 
\STATE $H \gets Hash(S)$
\STATE $Z \gets  \floor*{\frac{H\cdot 10000}{float(0xFFFFFFFFFFFFFFFF)}}$
  \IF{$\frac{Z}{100} >= E\% \cdot 100$}
    \STATE \RETURN ignore
  \ENDIF
 \IF{$Z\mod100 < C\% \cdot 100$}
    \STATE \RETURN control
  \ELSE
    \STATE \RETURN treatment
  \ENDIF
\end{algorithmic}
\end{algorithm} 
By design, this algorithm guarantees consistent assignment since it is a deterministic mapping from concatenated experiment ID and user ID to an assigned bucket. Further, the algorithm leads to a monotonic ramp-up by comparing $Z/100$ with $E\%$, and only users who get exposed are eligible to receive bucket assignments. 

The choice of hash functions affects the speed and the outcome of the algorithm. We evaluated three hash functions with the new algorithm. The three hash functions are FNV, MD5, and SpookyHash \citep{jenkins2012spookyhash}. We refer to the original randomization algorithm, the new algorithm with FNV hash, the new algorithm with MD5 hash, and the new algorithm with SpookyHash as algo 1, algo2, algo3, and algo 4 respectively in the remaining of the paper.
\subsection{Statistical Evaluations}

We conduct Chi-square test \citep{pearson1900x} to verify the uniformity of distribution of $R_b|R_e$ at various values of $R_e$. Specifically, after fixing the value $R_e = y$, we run the Chi-square goodness of fit test with the null hypothesis that $R_b|R_e = y$ follows a uniform distribution.
\begin{equation*}
f(R_b=x|R_e = y)=\frac{1}{100}  \quad x=0,1,\cdots,99;\hspace{.1cm} y=0,1,\cdots,99
\end{equation*}

In addition, we conduct a Chi-square test to evaluate the independence between $R^i_b$ and $R^j_b$ from experiment $i$ and experiment $j$. That is, we set both $R_e^i$ and and $R_e^j$ to 100\% and then apply Chi-square Independence test to $R^i_b$ and $R^j_b$ with the null hypothesis that $R_b^{i} \indep R_b^{j}$.

\section{Results}

\subsection{Uniformity test}

The new algorithm yields a uniform distribution of $R_b|R_e<y$ for all $y$ regardless of the choice of the hash function. In contrast, the original algorithm leads to non-uniform distribution of $R_b|R_e<y$ when $y\neq 100$.

\begin{figure}[htbp]
    \centering
    \includegraphics[width=10cm]{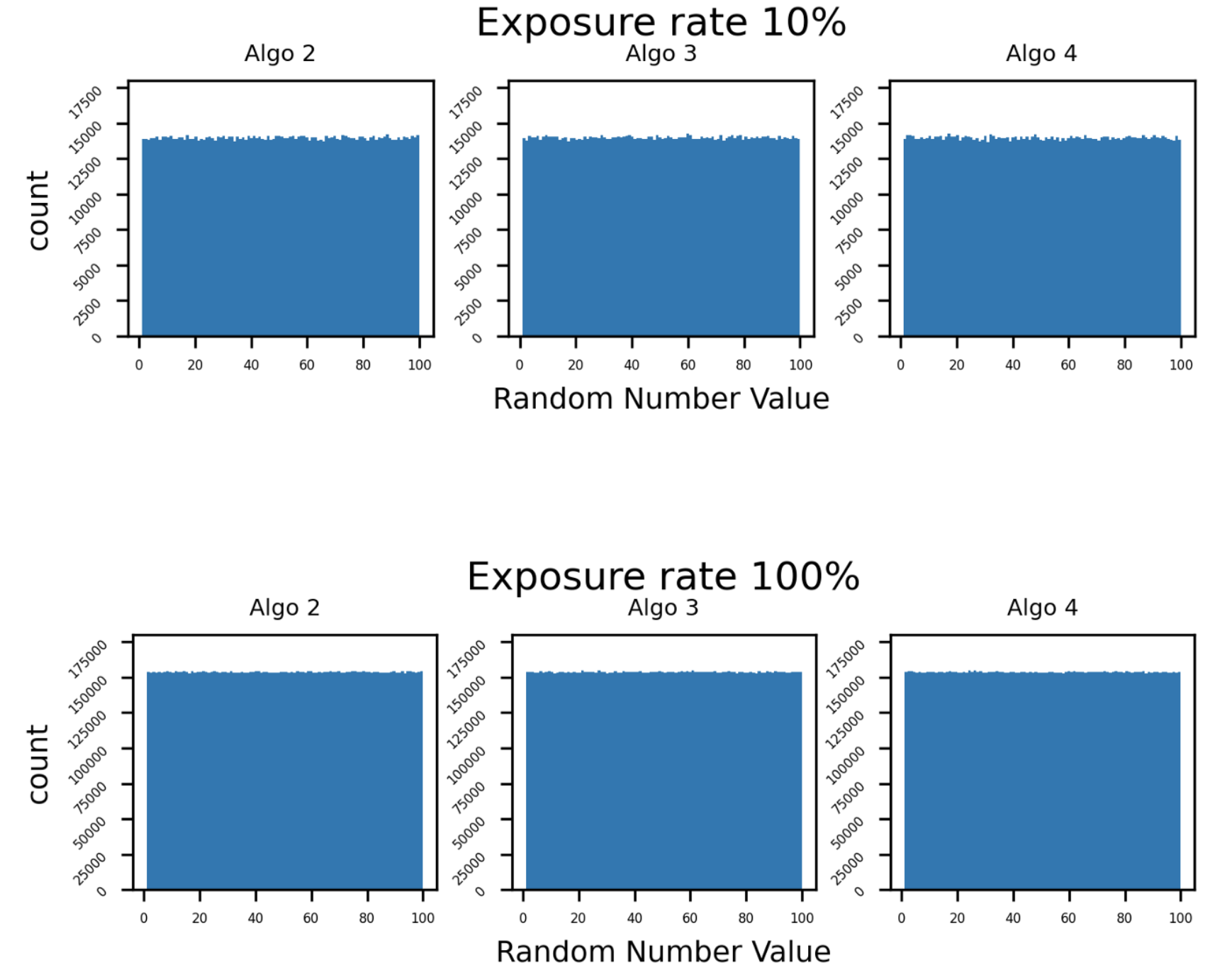}
    \caption{Historgram of $R_b|R_e<10$ (top row) and $R_b|R_e<100$ (bottom row) resulted from the new algorithm}
    \label{fig:uniform-test}
\end{figure}

Figure \ref{fig:uniform-test} demonstrates the uniform distribution of $R_b|R_e<100$ and $R_b|R_e<10$ resulted from the new algorithm with three hash functions respectively. The distributions of $R_b|R_e<y$ also look uniform for other values of y (data not shown). On the contrary, when using the old algorithm, $R_b|R_e<y$ look uniform only when $y=100$ (Figure \ref{fig:uniform}). Data is not shown for other values of y except 10 and 100. 

The results of Chi-square goodness of fit tests align with the histograms in Figure \ref{fig:uniform} and Figure \ref{fig:uniform-test}. Table \ref{table:uniform} presents the results of the Chi-square goodness of fit tests conducted for the original algorithm and the new algorithm with three hash functions, respectively. When \emph{p}-value is below 0.05, it indicates that the distribution of $R_b|R_e<y$ significantly deviates from uniform distribution. The \emph{p}-values are above 0.05 for the new algorithm with exposure rates 100\%, 10\%, and other exposure rates (data not shown). Contrarily, the \emph{p}-values are highly significant for the original algorithm with 10\% exposure rate and other exposure rates (data not shown) except the exposure rate 100\%.

\begin{table}[h]
\footnotesize
\caption{Results of Chi-square goodness of fit test}
\centering
%     \begin{adjustbox}{width=\textwidth,center}
    % \begin{adjustbox}{center}
        \begin{tabular}{llllll}
            \hline
            && \bf \makecell{Original\\Algorithm} 
            & \bf \makecell{New Algorithm \\(FNV)}
            & \bf \makecell{New Algorithm \\(MD5)}
            & \bf \makecell{New Algorithm \\(SpookyHash)}
            \\ 
            \hline
            \multirow{2}{*}{\shortstack{10\%\\ exposure}} & \multicolumn{1}{l|}{Test Statistic} 
            & \multicolumn{1}{c}{62627.92} 
            & \multicolumn{1}{c}{85.27} 
            & \multicolumn{1}{c}{76.80} 
            & \multicolumn{1}{c}{106.00} 
            \\
            \cline{2-6} 
            & \multicolumn{1}{l|}{p-value} 
            & \multicolumn{1}{c}{0.0000}
            & \multicolumn{1}{c}{0.8356} 
            & \multicolumn{1}{c}{0.9521} 
            & \multicolumn{1}{c}{0.2968} 
            \\
            \cline{2-6}
            \multirow{2}{*}{\shortstack{100\%\\ exposure}} & \multicolumn{1}{l|}{Test Statistic} 
            & \multicolumn{1}{c}{106.74} 
            & \multicolumn{1}{c}{113.22} 
            & \multicolumn{1}{c}{108.19} 
            & \multicolumn{1}{c}{108.40} 
            \\
            \cline{2-6} & \multicolumn{1}{l|}{p-value} 
            & \multicolumn{1}{c}{0.2798}
            & \multicolumn{1}{c}{0.1556} 
            & \multicolumn{1}{c}{0.2480} 
            & \multicolumn{1}{c}{0.2435} 
            \\
            \hline
        \end{tabular}
%     \end{adjustbox}
%     \vspace{ - 05 mm}
\label{table:uniform}
\end{table}

\subsection{Independence test}\label{sec:ind}
The new algorithm leads to independent assignments with hash functions SpookyHash and MD5 but not with FNV. The old algorithm yields dependent assignments.

\begin{figure}[htbp]
    \centering
    \includegraphics[width=10cm]{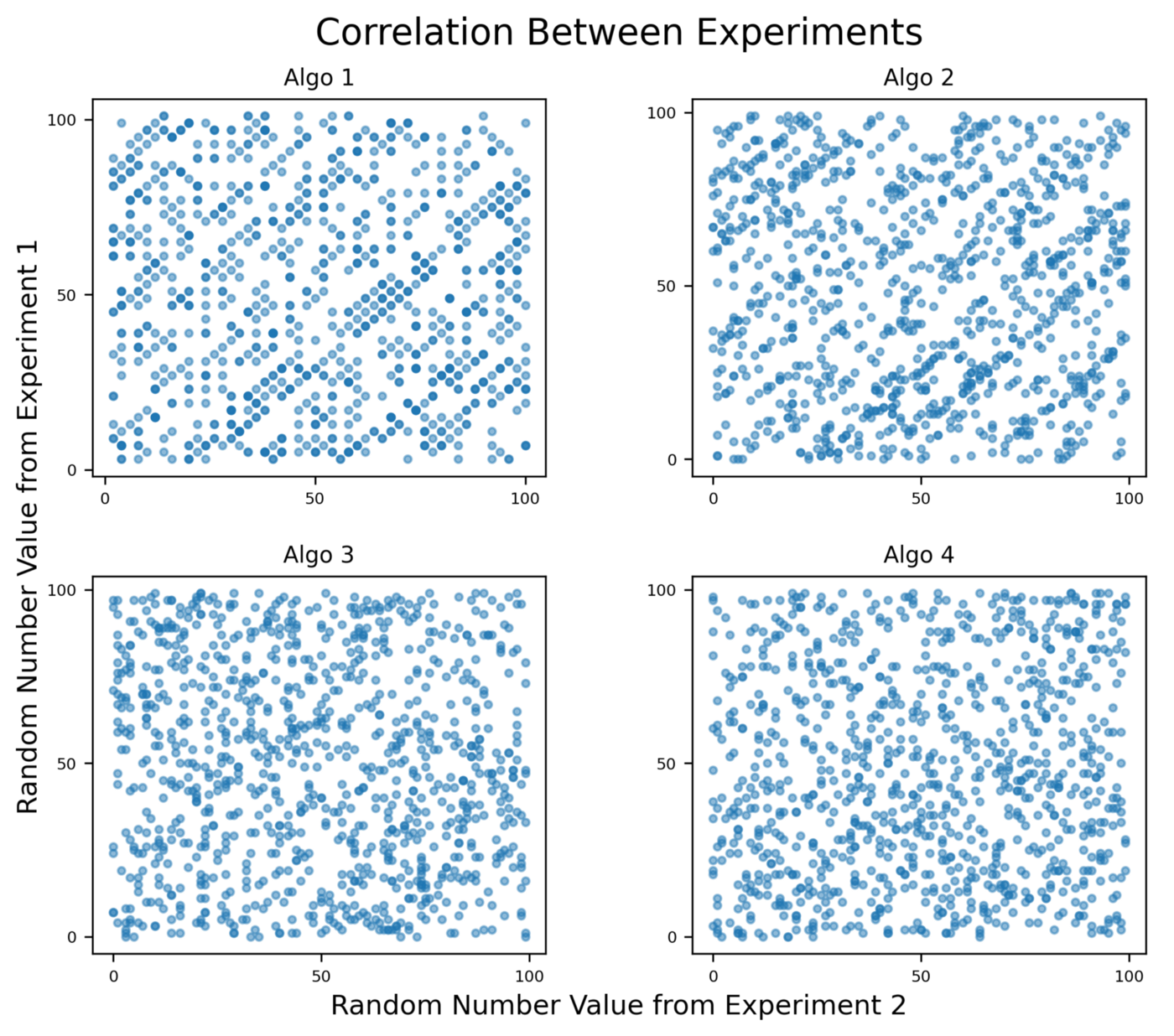}
    \caption{Scatter plots of $R_b|R_e<100$ values in one experiment vs them in another experiment. Each panel illustrates $R_b|R_e<100$ values of 1000 randomly selected users.}
    \label{fig:independence}
\end{figure}

Figure \ref{fig:independence} demonstrates the correlation between the assignments from two experiments. We randomly selected one thousand users and plotted their values of $R_b|R_e<100$ from two experiments. The scatter plots corresponding to the original algorithm (top left) and the new algorithm with FNV (top right) demonstrate the correlation between the assignments of the two experiments. In contrast, the scatter plots of the new algorithm with MD5 (bottom left) and the new algorithm with SpookHash (bottom right) both look random.  

The results of the Chi-square independence tests confirm the dependent assignments yielded by the original algorithm and the new algorithm with FNV, and the independent assignments resulted from the new algorithm with MD5 and SpookyHash.
Table \ref{table:ind} reports the results of the Chi-square independence test conducted for both the new and the original algorithms. The \emph{p}-values are highly significant for the original algorithm and the new algorithm with FNV. This finding suggests FNV hash function is not suitable for assignment, which aligns with conclusion from \cite{zhao2016online}. When using the new algorithm with MD5 and SpookyHash, the \emph{p}-values are both above 0.05. 
\begin{table}[h]
\caption{Chi-square independence test}
\centering
%     \begin{adjustbox}{width=\textwidth,center}
    % \begin{adjustbox}{center}
        \begin{tabular}{lllll}
            \hline
            & \bf \makecell{Original\\  Algorithm} & \bf \makecell{New Algorithm \\ (FNV)}  & \bf \makecell{New Algorithm \\ (MD5)} & \bf \makecell{New Algorithm \\  (SpookyHash) }\\ 
            \hline
             \multicolumn{1}{l|}{Test Statistic} 
             & \multicolumn{1}{c}{65883.26} 
             & \multicolumn{1}{c}{13366.33} 
             & \multicolumn{1}{c}{0.44} 
             & \multicolumn{1}{c}{0.31} 
             \\\cline{1-5}
            \multicolumn{1}{l|}{p-value} 
            & \multicolumn{1}{c}{0.0000}
            & \multicolumn{1}{c}{0.0000}
            & \multicolumn{1}{c}{0.5075}
            & \multicolumn{1}{c}{0.5783} \\
            \hline

        \end{tabular}
%     \end{adjustbox}
%     \vspace{ - 05 mm}
\label{table:ind}
\end{table}

\subsection{Computation Speed}
The experimentation service at Wish has hundreds of thousands of queries per second (QPS), and the QPS grows quickly in a hyper-growth company. Therefore, experiment bucket evaluation latency is critical. In our latency testing, we found that the new algorithm is about four times faster than the original algorithm.

\section{Discussions}
Although controlled experimentation has been well studied, there are unique challenges when applying online controlled experimentation (aka A/B testing) at scale. Seemingly simple randomization can be hard to get right. Yet, good randomization is critical to establishing causal conclusions. Our improved randomization algorithm satisfies the statistical requirements and is faster.

\section{Acknowledgement}

We want to thank Chao Qi, Shawn Song, Lance Deng, Caroline Davey, Gus Silva, Iryna Shvydchenko for insightful discussions and contributions to the implementation.

%Note:BibTeX also works
\bibliographystyle{apalike}
\bibliography{references}  

\end{document}